\documentstyle[aps,prl,twocolumn,floats,epsfig]{revtex} 
\begin{document}
\draft
 \wideabs{     
\title{The Hamiltonian of the V$_{15}$ Spin System
from first-principles Density-Functional Calculations}
\author{Jens Kortus\cite{MPI}, C. Stephen Hellberg and Mark R. Pederson}
\address{Center for Computational Materials Science,
Code 6390, Naval Research Laboratory, Washington, DC 20375}
\date{\today}
\maketitle
\begin{abstract}
We report first-principles all-electron density-functional based
studies of the electronic structure, magnetic ordering and anisotropy
for the V$_{15}$ molecular magnet. 
From these calculations, we determine
a Heisenberg Hamiltonian
with five antiferromagnetic and one {\em ferromagnetic} exchange
couplings.  We perform
direct diagonalization to determine the temperature dependence of the 
susceptibility.
This Hamiltonian reproduces the experimentally observed spin
$S$=1/2 ground state and low-lying $S$=3/2 excited state.
A small anisotropy term is necessary
to account for the temperature independent
part of the magnetization curve. 

\end{abstract}   
}

With the continued interest in the fabrication and optimization of 
miniaturized magnetic devices\cite{sci-nano}
future design considerations will require
an understanding of nanoscale magnetic systems.  In order to transition
such materials into simple devices it is necessary to be able to explain how
interactions such as spin-spatial coupling and spin-spin
exchange effects may couple collectively to create a seemingly 
single-spin system.  Further, it is necessary to determine the temperature 
range at which such systems will indeed behave collectively. 
In general the properties of
a nanoscale system of coupled spins depend directly on the strength 
of the exchange-parameters and on the size and sign of the anisotropy 
energy due to spin-orbit coupling. 
While these parameters are generally determined by the transition metal 
atoms, the ligands and other nonmagnetic atoms are responsible for 
stabilizing the array of spins.
Requisite to a complete computational 
understanding of such spin systems is the ability to account for the
strong ligand-metal interactions and to determine whether the behavior
of a given spin system is mainly mediated by the anisotropy, by spin-spin
coupling or by a combination of the two. 

Recently, the Mn$_{12}$-Acetate and Fe$_{8}$ molecules\cite{Mn12}
have attracted considerable interest because they behave as 
high-single-spin systems (total spin $S$=10)
at temperature ranges on the order of 20-60K. 
Due to their large magneto-molecular anisotropy energy these systems 
retain their moment orientation at reasonably
high temperatures and exhibit the phenomena of resonant tunneling of
magnetization at well defined magnetic fields\cite{Fried,Sangre}.

The $ \rm K_6[V_{15}As_6O_{42}(H_2O)]~8H_2O $ 
molecular crystal, first
synthesized by M{\"u}ller and D{\"o}ring\cite{exp1,nature}, represents
a transition-metal spin system in the same size regime as the Mn$_{12}$ and
Fe$_{8}$ molecular crystals.
In contrast to Mn$_{12}$ and Fe$_{8}$  molecules, the V$_{15}$ molecule
is thought to behave as a weakly anisotropic magnet composed of 
15 spin $s$=1/2 particles which couple together to form a 
molecule with a total spin $S$=1/2 ground state.  
Besides the fundamental interest in understanding quantum effects
in these nanomagnets they might be also relevant for 
implementations of quantum computers\cite{merm}.
Calculations on such 
correlated systems present a challenge to mean-field frameworks such 
as density-functional theory because it is often not possible to construct 
a single collinear reference state which preserves the inherent symmetry 
of the system and has the correct spin quantum numbers. 

This work utilizes an efficient coupled multilevel analysis which relies
on fitting density-functional energies to 
mean-field Heisenberg or Ising energies 
in order to determine the exchange parameters.
The approximate exchange parameters gleaned from the first $N$ 
Ising configurations were used to find the next
lowest energy Ising configuration and subsequently
to improve the parameterization of the exchange parameters. 
 ``Self Consistency'' of this approach is determined 
when the predicted Ising levels are unchanged by the
addition of data from new Ising configurations.
The coupling of the density-functional method to a
classical Ising representation allowed us to determine the exchange 
parameters by considering only several spin configurations.  
Further, the resulting
ground-state spin configuration within density-functional theory exhibits
the correct spin projection of $1/2$.
With the exchange parameters determined, we diagonalize the 
complete many-body Heisenberg Hamiltonian to calculate the
susceptibility and spin correlation functions for comparison
with experiment.
The many-body basis is complete, so all states are allowed
including non-collinear spin arrangements and quantum disordered phases
\cite{hell1}.

Starting from X-ray data\cite{exp2} we generated several 
unit cells and isolated a single $ \rm K_6[V_{15}As_6O_{42}(H_2O)] $ unit.
In order to optimize the geometry within the quasi-$D_3$ symmetry of the 
V$_{15}$ molecule\cite{exp1}, we initially replaced 
the statistically oriented 
water molecule at the center of the molecule by a neon atom and used
an $S$=3/2 spin configuration which does not break the crystallographic 
symmetry.  The Ising configuration of this molecule consists of three aligned 
spin-$\frac{1}{2}$ V atoms in the triangle and  equivalent upper 
and lower hexagons composed of a ring of antiferromagnetically (AF) coupled 
spin-$\frac{1}{2}$ V atoms. 

\begin{figure}              
\begin{center}
\psfig{file=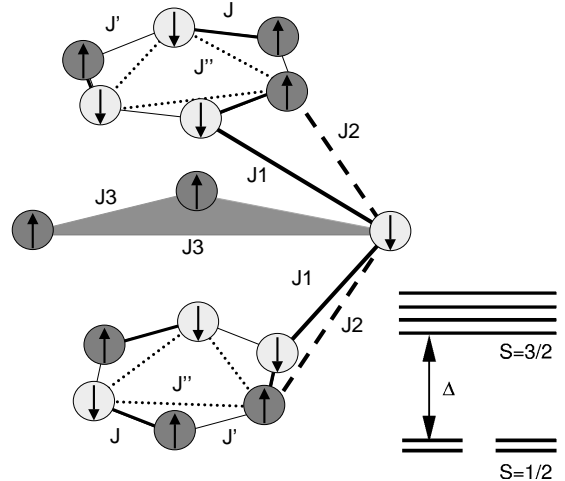,width=2.5in,angle=-90,clip=true}
\end{center}
\caption{
The 15 magnetic vanadium atoms 
of the $ \rm K_6[V_{15}As_6O_{42}(H_2O)] $ molecule. 
They form two hexagonal layers and
an inner triangular layer of vanadium atoms sandwiched within.
The arrows show the lowest energy spin configuration found
from DFT calculations.  The six exchange parameters 
used in the Heisenberg Hamiltonian are shown as lines.
Schematically displayed are energy levels of the 
Kramer doublet ($S$=1/2)
ground state and the low lying quadruplet ($S$=3/2)
separated by $\Delta$.
}
\label{fig1}
\end{figure}

The geometry of the molecule was then optimized within
the all-electron density-functional methodology
using the generalized-gradient approximation (GGA)\cite{pbe}. 
The calculations were performed with the
Naval Research Laboratory Molecular Orbital Library (NRLMOL)\cite{codes}.  
Calculations on 49 geometrical configurations were 
performed during the conjugate-gradient relaxation of the molecule.
Subsequent calculations show that the geometrical, electronic and magnetic 
properties of this system are unaffected by the presence or type of inert 
moiety enclosed within the void.  
Using this geometry, we performed eleven additional calculations on 
different spin configurations (See Table \ref{tab1}) to 
determine the six exchange parameters ($J'$s) of the Heisenberg Hamiltonian
\begin{equation}
H=\sum J_{ij} {\bf S}_i\cdot {\bf S}_j,
\label{hamiltonian}
\end{equation}  
as well as the  spin configuration of the density-functional 
ground state\cite{hell1}.
The $J'$s used in the above Hamiltonian
are defined according to Fig.~\ref{fig1}. As shown in Table \ref{tab1}, we 
have included high-spin configurations (XI, XII, XIII), which generally 
have some symmetry as well as lower-spin non-symmetric configurations.
The energy for the high-spin $S=15/2$ ferromagnetic (FM) state (XIII) of 
873~meV is predominantly caused by a large AF 
exchange coupling ($J$) between the most closely bonded hexagonal V atoms.  
However, the 113~meV splitting between the 
 $S$=9/2 and $S$=15/2 states (XII and XIII)    
shows that there is a reasonably strong AF coupling,
approximately 18~meV on average, between the triangular and hexagonal 
atoms.  All of the data displayed in
Table \ref{tab1} has been used to determine the exchange parameters from a
least square fit to the mean-field solution of the Heisenberg 
Hamiltonian (\ref{hamiltonian}).
The fit is very good with errors ranging from 
0.1-1.55 meV.  The fit leads to exchange
parameters of $J$ = 290.3~meV, $J'$= -22.7~meV, $J''$=15.9~meV, 
$J1$ = 13.8~meV, $J2$ =  23.4~meV and $J3$ = 0.55~meV, 
where positive numbers correspond to AF and negative ones to FM
interactions.
The {\em ferromagnetic} interaction $J'$ is a 
surprising result \cite{Korotin} and deserves further discussion since it is 
qualitatively different from earlier assumptions based on
entirely AF interactions\cite{nature,chio}. 
Ferromagnetic coupling is possible without polarizing the oxygens 
through a 4'th order process similar to super-exchange.
In super-exchange, the intermediate state has the lowest
$d$-orbital on the V doubly occupied with up and down electrons\cite{gooden}.
However, electrons can also hop to higher energy
$d$-orbitals on the V's. 
In this case both parallel and antiparallel spins are allowed
without violating the Pauli exclusion principle,
but Hund's rule coupling prefers parallel alignment.
The super-exchange process (same $d$-orbital) completely excludes
the process with same-spin electrons while the FM
process (different $d$-orbitals) merely favors FM
alignment.  Thus a FM coupling is
obtained if the V-O hopping matrix elements into the higher
$d$-orbital are significantly larger than the matrix
elements for the hopping of O electrons into the lowest energy
$d$-orbital.  
The occurrence of such interactions 
are possible in a low-symmetry system such as the one studied here.

Even with this FM interaction, our spin Hamiltonian yields
an $S$=1/2 ground state composed largely of Ising configurations
similar to the one depicted in Fig.~\ref{fig1}.
This Ising configuration was predicted from the $J$'s
from earlier fits to DFT energies and corresponds 
to the ground state DFT configuration (I).

\begin{table}[bh!]
\caption{DFT
energies ($E$) of calculated
Ising configurations, energies obtained from the fit,
and $4\langle S^q_i S^q_j \rangle$ along each of the six bonds.
Also included 
is the anisotropy shift $\delta$ for the $M_s=S$ state of each Ising 
configuration. A least square fit of this data leads to exchange
parameters of $J$=290.3, $J'$=-22.7, $J$''=15.9, $J1$=13.8,
$J2$=23.4 and $J3$=0.55~meV.
}
\label{tab1}
\begin{tabular}{ddrrrrrrrrc}
$E$ (meV)&Fit & $J$& $J'$& $J''$&$J1$& $J2$&$J3$&  Spin& Label& $\delta$ (K)\\
\hline
-78.37 & -78.44 &  -6  &  2 & -2 &  6 & -6 & -1 & 1/2 &I    & 0.8 \\
-73.39 & -73.63 &  -6  &  2 & -2 &  4 & -4 & -1 & 1/2 &II   &     \\
-35.48 & -35.08 &  -6  & -2 &  2 &  4 & -4 & -1 & 1/2 &III  &     \\
-34.89 & -34.53 &  -6  & -2 &  2 &  4 & -4 &  3 & 3/2 &IV   &     \\
  0.00 &  -0.79 &  -6  & -6 &  6 &  6 & -6 &  3 & 3/2 &V    & 1.5 \\
  8.38 &   8.28 &  -6  & -6 &  6 &  2 & -2 & -1 & 1/2 &VI   & 1.3 \\
 28.14 &  28.08 &  -6  & -6 &  6 & -6 &  6 &  3 & 3/2 &VII  &     \\
126.32 & 126.14 &  -4  & -4 &  6 &  4 & -6 &  3 & 1/2 &VIII &     \\
129.17 & 128.88 &  -4  & -4 &  2 &  6 & -4 &  3 & 5/2 &IX   &     \\
278.35 & 278.50 &  -2  & -6 &  2 &  4 & -4 &  3 & 3/2 &X    &     \\
434.22 & 435.78 &   0  &  0 &  6 &  6 &  0 &  3 & 9/2 &XI   & 1.6 \\
760.75 & 760.76 &   6  &  6 &  6 & -6 & -6 &  3 & 9/2 &XII  & 1.6 \\
873.11 & 872.35 &   6  &  6 &  6 &  6 &  6 &  3 &15/2 &XIII & 1.8 \\
\end{tabular}
\end{table}

We have fully diagonalized the Heisenberg Hamiltonian (\ref{hamiltonian}).  
Using all symmetries, the largest irreducible many-body
subspace has dimension 2145.  We find a spin-1/2 Kramer
doublet as the ground state with a low-lying spin-3/2
quadruplet as shown in Fig.~\ref{fig1}.
The rest of the spectrum is well separated from these eight states.
The large value of $J$ binds the spins in the hexagons into singlets.
The low-energy physics arises from the inner triangle spins interacting
with each other both directly and 
with an effective coupling through the hexagons,
yielding the doublet-quadruplet spectrum \cite{dobro}.
There are two important energy scales in the spectrum:
$\Delta$, the gap between the doublet and quadruplet, and $J$,
the energy at which the singlets in the hexagon break
and the molecule starts to behave as more than three spins.

The low-energy effective interaction between inner triangle spins
proceeds via $J1$ and $J2$ (which frustrate each other)
to a hexagon singlet, via $J'$ and $J''$ to a neighboring singlet,
and finally via $J1$ and $J2$.
A larger contribution to $\Delta$ comes from a direct interaction
through $J3$ mediated by hopping through O and As levels.
Thus simple perturbation theory \cite{nature} yields 
\begin{equation}
\Delta = \frac{3}{4} \frac{(J2-J1)^2(J''-J')}{J^2} + \frac{3}{2} J3.
\label{perturb}
\end{equation}

Comparing our calculated susceptibility with experiment \cite{chio},
we find the low-temperature behavior indicates our doublet-quadruplet gap
$\Delta \approx 10$K to be significantly larger than the experimental
value of $\Delta \approx 3.7$K,
while the high-temperature behavior shows our calculated value of $J$ 
is too large.
Both of these discrepancies 
can be explained almost entirely by  
$J'$s that are too large within the density-functional-based treatment,
as known for other vanadium systems \cite{hell1}.

\begin{figure}
\begin{center}
\psfig{file=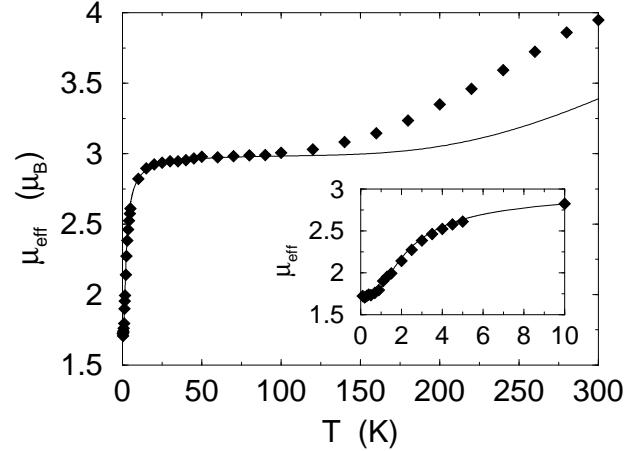,width=3.2in,clip=true}
\end{center}
\caption{The effective moment $\mu_{\rm eff} = \sqrt{3 \chi T}$
calculated with all $J'$s uniformly rescaled by dividing by 
a constant factor of 2.9.
The diamonds are the actual experimental values from Chiorescu.
The inset shows the low temperature behavior.}
\label{fig2}
\end{figure}

Agreement with experiment for the low temperature
$\Delta$ can be achieved by dividing all $J'$s by a
constant factor of 2.9. Fig.~\ref{fig2}
shows our calculated result with rescaled $J'$s
compared to experimental data from Chiorescu.
A uniform scaling of our calculated exchange parameters is 
not able to obtain the right low and high temperature
behavior at the same time. 
The high temperature behavior could be improved by futher reducing
$J$ with corresponding
adjustments to the other couplings to keep $\Delta$ constant.

A set of AF interactions \cite{nature} also fits the experimental results.
In fact any set of parameters with the correct values
of $J$ and $\Delta$ given by Eq.\ (\ref{perturb})
will fit the effective moment well.
To support future efforts aimed at experimentally confirming
our relative parameters, we have calculated the
spin-spin correlation function
$C_S = 4\langle \boldmath{S}_i^q\cdot \boldmath{S}_j^q \rangle$,
with $q$ representing the arbitrary quantization axis.
Since $J$ is always largest,
the spins tend to form a singlet along this bond,
yielding $C_S(J)\cong -1$ (AF correlation).
The three inner triangle spins form a non-collinear arrangement
with $C_S \cong -0.33$ between all pairs of spins.
The direct $J3$ coupling reduces the correlations between triangle
and hexagon spins.
We find 
$C_S(J1)\cong+0.02$
and 
$C_S(J2)\cong-0.02$
(FM and AF correlations, respectively),
while the coupling constants of Ref.~\cite{nature} yield
$C_S(J1)\cong+0.12$
and 
$C_S(J2)\cong-0.14$.
This prediction should be measurable with neutron scattering.

Chiorescu {\em et al.}\ observe that rotation of the spin projection 
is achieved without encountering a barrier\cite{chio}
and Dobrovitski {\em et al.}\ posit that the V$_{15}$ molecule is indeed a 
low anisotropy system\cite{dobro}.  
As shown below the existence of either 
easy-plane or easy-axis anisotropy will shift the 
$M_s$=3/2-1/2 Zener-Landau tunneling transitions 
that have been observed by Chiorescu {\em et al.}\cite{chio}.

Recently, Pederson and Khanna have developed a new method for accounting
for second-order anisotropy energies\cite{ped-so}.
This method relies on a simple
albeit exact method for spin-orbit coupling and a second-order perturbative
treatment of the spin-orbit operator to determine the dependence of
the total energy on spin projection. Initial applications to Mn$_{12}$
lead to a density-functional-based second-order anisotropy energy of
55.7K\cite{ped1} 
which is in essential agreement with the experimentally deduced 
values of 55.6K\cite{barra}.  
We have generalized this methodology to systems with
arbitrary symmetry and have calculated the anisotropy energy for several
different spin configurations of the V$_{15}$ molecule.

We have calculated the anisotropy energy for the lowest Ising 
configurations with one, three, nine and fifteen unpaired 
electrons, as given in Table \ref{tab1}. 
In all cases we find that the V$_{15}$ possesses weak easy-plane anisotropy.
This result ensures that anisotropy effects will not change the total
spin of the V$_{15}$ ground state.  Examination of the expression for the
second order anisotropy energy in Ref. \cite{ped-so}, 
shows that such energies do
not necessarily scale as the square of the total moment.
Indeed, as shown in Table \ref{tab1},
we find that the anisotropic effects are in fact only weakly 
dependent on  the total spin and that the energy of 
the M$_s$=S state increases by approximately $\delta$=0.8K to 1.8K. 
Chiorescu {\em et al.} show that the broadening of the Zener-Landau
tunneling fields decreases with temperature 
for the $|1/2,1/2\rangle$ to $|1/2,-1/2\rangle$
transitions but are independent of temperature for the 
$|3/2,3/2\rangle $ to the $|3/2,1/2\rangle $ transitions\cite{chio}.  
This behavior is exactly what is expected 
from a sample containing weak easy-plane spin anisotropy.  
At sufficiently low
temperatures, 
only the $S=3/2$ and $S=1/2$ states 
are relevant, and the field-dependent crossing of these states 
depends on whether the 
magnetic field is parallel or perpendicular to the easy-plane.
The broadening is proportional to the difference
of the magnetic anisotropy energy for the different spin 
configurations involved. 
Due to the small anisotropy in V$_{15}$ the effect will be small.
Although it is not possible to translate the DFT 
obtained anisotropy energies directly to the quantum mechanical 
many-spin ground state discussed here,
we obtain from these energies  tunnel field broadenings between 0.1~T 
to 0.48~T which envelope the experimentally observed field broadening 
of about 0.2~T\cite{chio}. 
In powdered samples, the small easy-plane anisotropy would lead to
a broadening in the tunneling field and in single crystals the effect 
would change the tunneling fields as a function of field 
orientations.

To summarize, we have performed accurate all-electron density-functional 
calculations on the V$_{15}$ cluster as a function of geometry and spin
configuration. By dynamically coupling the mean-field density-functional 
approach to exact-diagonalization of a many-spin Heisenberg representation,
we have efficiently determined the lowest density-functional configurations
and the entire Heisenberg
spin excitation spectrum. Our calculations suggest that the
small experimentally observed orientational dependence of the tunneling 
field for the $M_s$=3/2 to $M_s$=1/2 is a signature of configuration 
dependent magnetic anisotropy in this molecule. 
The method used here is general and allows one to 
characterize both systems which are potentially useful for
magnetic storage (Mn$_{12}$, Fe$_8$) and systems which
show quantum coherence, such as the one studied here.

This work was supported in part by ONR grant N0001498WX20709
and N0001400AF00002.
Computations were performed at the DoD Major Shared Resource Centers.
We thank I. Chiorescu for enlightening discussions and for supplying
the data used in Fig.~\ref{fig2}.

\end{document}